\documentclass[aps,prl,twocolumn,superscriptaddress,showpacs,amsmath,amssymb,longbibliography]{revtex4-2}

\usepackage{graphicx}
\usepackage{dcolumn}
\usepackage{epsfig}
\usepackage{subfigure}
\usepackage{amssymb,amsmath}
\usepackage{epstopdf}
\usepackage{epsfig}
\usepackage{color}
\usepackage{amsfonts}
\usepackage{amscd}
\usepackage{float}
\usepackage{bm}
\usepackage[colorlinks,urlcolor=blue,linkcolor=blue,citecolor=blue]{hyperref}
\usepackage{bbm}
\usepackage{mathrsfs}
\usepackage{mathtools}
\usepackage{diagbox}
\usepackage[dvipsnames]{xcolor}
\usepackage{ulem}

\definecolor{zhou}{RGB}{255 130 71}
\newcommand{\orcid}[1]{\href{https://orcid.org/#1}{\includegraphics[width=6.6pt]{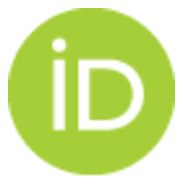}}}

\begin{document}

\preprint{APS/123-QED}

\title{Approximate Autonomous Quantum Error Correction with Reinforcement Learning}
\author{Yexiong Zeng\orcid{0000-0001-9165-3995}}
\affiliation{Theoretical Quantum Physics Laboratory, Cluster for Pioneering Research, RIKEN, Wakoshi, Saitama 351-0198, Japan}
\affiliation{Quantum Computing Center, RIKEN, Wakoshi, Saitama 351-0198, Japan}
 
\author{Zheng-Yang Zhou}%
\affiliation{Theoretical Quantum Physics Laboratory, Cluster for Pioneering Research, RIKEN, Wakoshi, Saitama 351-0198, Japan}%

\author{Enrico Rinaldi\orcid{0000-0003-4134-809X}}
\affiliation{Quantinuum K.K., Otemachi Financial City Grand Cube 3F, 1-9-2 Otemachi, Chiyoda-ku, Tokyo, Japan}
\affiliation{Theoretical Quantum Physics Laboratory, Cluster for Pioneering Research, RIKEN, Wakoshi, Saitama 351-0198, Japan}
\affiliation{Quantum Computing Center, RIKEN, Wakoshi, Saitama 351-0198, Japan}
\affiliation{Department of Physics, University of Michigan, Ann Arbor, Michigan, 48109-1040, USA}
\affiliation{Interdisciplinary Theoretical and Mathematical Sciences Program (iTHEMS), RIKEN, Wakoshi, Saitama 351-0198,
	Japan}

\author{Clemens Gneiting\orcid{0000-0001-9686-9277}}
\altaffiliation[clemens.gneiting@riken.jp]{}
\affiliation{Theoretical Quantum Physics Laboratory, Cluster for Pioneering Research, RIKEN, Wakoshi, Saitama 351-0198, Japan}
\affiliation{Quantum Computing Center, RIKEN, Wakoshi, Saitama 351-0198, Japan}

\author{Franco Nori\orcid{0000-0003-3682-7432}}
\altaffiliation[fnori@riken.jp]{}
\affiliation{Theoretical Quantum Physics Laboratory, Cluster for Pioneering Research, RIKEN, Wakoshi, Saitama 351-0198, Japan}
\affiliation{Quantum Computing Center, RIKEN, Wakoshi, Saitama 351-0198, Japan}
\affiliation{Department of Physics, University of Michigan, Ann Arbor, Michigan, 48109-1040, USA}


\begin{abstract}
Autonomous quantum error correction (AQEC) protects logical qubits by engineered dissipation and thus circumvents the necessity of frequent, error-prone measurement-feedback loops. Bosonic code spaces, where single-photon loss represents the dominant source of error, are promising candidates for AQEC due to their flexibility and controllability. While existing proposals have demonstrated the in-principle feasibility of AQEC with bosonic code spaces, these schemes are typically based on the exact implementation of the Knill-Laflamme conditions and thus require the realization of Hamiltonian distances $d\geq 2$. Implementing such Hamiltonian distances requires multiple nonlinear interactions and control fields, rendering these schemes experimentally challenging. Here, we propose a bosonic code for approximate AQEC by relaxing the Knill-Laflamme conditions. Using reinforcement learning (RL), we identify the optimal bosonic set of codewords (denoted here by RL code), which, surprisingly, is composed of the Fock states $\vert 2\rangle$ and $\vert 4\rangle$. As we show, the RL code, despite its approximate nature, successfully suppresses single-photon loss, reducing it to an effective dephasing process that well surpasses the break-even threshold. It may thus provide a valuable building block toward full error protection. The error-correcting Hamiltonian, which includes ancilla systems that emulate the engineered dissipation, is entirely based on the Hamiltonian distance $d=1$, significantly reducing model complexity. Single-qubit gates are implemented in the RL code with a maximum distance $d_g=2$.
\end{abstract}

\maketitle


{\it Introduction.}---Implementing {efficient} quantum error correction (QEC) is a prerequisite and the main obstacle {toward} building a general-purpose quantum computer \cite{Chiaverini2004Dec, Schindler2011May, RevModPhys.87.307}. The purpose of QEC is to {restore} encoded quantum information that has been {corrupted by environmental noise:} physical qubits inevitably interact with the surrounding environment \cite{Gaitan2017Jan,lidar_brun_2013}.~In conventional QEC {this is achieved by} error syndrome measurements and adaptive recovery operations, which {involve} imperfect measurements and classical feedback loops that {themselves} raise and propagate errors \cite{PhysRevX.11.041058,PhysRevA.72.012306, PhysRevA.90.062344, Albert2019Jun, Ma2020Aug, PhysRevX.3.041013, PhysRevA.103.042406,PhysRevApplied.18.024076,PhysRevResearch.3.033275}.

Autonomous quantum error correction (AQEC) {has been developed as an alternative that avoids these additional error sources by realizing} QEC through quantum reservoir engineering, {where the cleverly designed and continuously acting dissipation processes replace the repeated measurement-feedback cycles} \cite{Gertler2021Feb, PhysRevLett.85.856, Perez2020Nov, PhysRevLett.111.120501, PhysRevLett.77.793}. {More specifically, the interaction between the to-be-protected system and an ancilla system is engineered to transport the decoherence-induced, cumulative entropy from the system to the ancilla, from where it decays into the environment \cite{PhysRevLett.116.150501, PhysRevLett.105.040502, Reiter2017Nov, Krastanov2021Jan, Cai2021Jan,Xu2022Oct,FernandoQ,Zanardi2016Feb,PhysRevA.98.012317}.}

{The realization of AQEC in bosonic systems is particularly attractive, as an infinitely large Hilbert space offers ample design opportunities while the noise channels remain fixed.}
Indeed, several pioneering works have demonstrated the great potential of AQEC to counteract single-photon loss, the dominant error source in bosonic systems; for instance, based on the T4C code \cite{Gertler2021Feb}, the binomial code \cite{Hu2019May}, or the $\sqrt{3}$ code \cite{PRXQuantum.3.020302}.~While these codewords fully satisfy the Knill-Laflamme (KL) conditions, {and hence in principle allow for} exact AQEC \cite{PhysRevA.55.900, PhysRevLett.84.2525}, {experimental limitations have so far prohibited their faithful implementation.~Therefore, the experimental surpassing of the break-even threshold (i.e, improved performance compared to codewords consisting of the Fock states $\vert 0\rangle$ and $\vert 1\rangle$ and no error correction) using AQEC is still lacking.}

{Rigorous implementation of the KL condition unavoidably implies codewords that entail complex and fragile superpositions of Fock states, which are hard to produce. This is reflected by the required Hamiltonian distance, i.e., the number of Fock states that must be bridged by the Hamiltonian. Indeed, an exhaustive search has excluded the existence of a code space that both rigorously satisfies the KL condition and is content with Hamiltonian distance $d=1$ for AQEC \cite{PRXQuantum.3.020302}. On the other hand, the larger the Hamiltonian distance, the more high-order interactions are required, accompanied by an increasing number of control fields. The experimental hardness is further aggravated by the necessity to implement gates, which typically require even larger Hamiltonian distances.}

{Here, we pursue a different strategy, following the spirit of approximate QEC \cite{Leung1997Oct, PhysRevLett.104.120501, PhysRevLett.107.080501, Kong2022Apr, PhysRevX.10.041018}. By relaxing the KL condition, we strive to ease the experimental overhead by lowering the Hamiltonian distance as much as possible. While this excludes the in-principle exact QEC, we consider our strategy successful, if the discovered AQEC surpasses the break-even threshold, i.e., if the encoded qubit outperforms the natural qubit and thus lowers the threshold for a concatenated QEC stack.}

{The search for the optimal approximate AQEC scheme that achieves this involves a complex optimization problem, which we solve with the help of reinforcement learning (RL) \cite{Sutton1998, PhysRevX.8.031086, Fosel2018Sep, PRXQuantum.2.040324, Guo2021Feb, PhysRevLett.127.190403,Bartkiewicz2020Jul,PhysRevLett.127.110502, PhysRevX.11.031070, Xiao2022Jan, PhysRevApplied.18.024033}. We find that the optimal AQEC scheme relies on surprisingly simple codewords (denoted here by RL code) composed of the two Fock states $\vert 2\rangle$ and $\vert 4\rangle$. The RL code not only significantly surpasses the break-even with an infidelity reduction of over 80\% (surprisingly even outperforming the $\sqrt{3}$ code), but also can be realized with the smallest possible Hamiltonian distance $d = 1$; the respective Hamiltonian distance required to implement single-qubit gates is $d_g=2$, which again outperforms all previously proposed code spaces.}

{We demonstrate that the approximate AQEC based on the RL code can be realized by complementing the encoded bosonic mode with an ancillary lossy mode, an ancillary lossy two-level system, and couplings that are readily available on existing platforms.}

{\it Approximate AQEC.}---
\begin{figure*}[htpb]
	\includegraphics[width=7.05in]{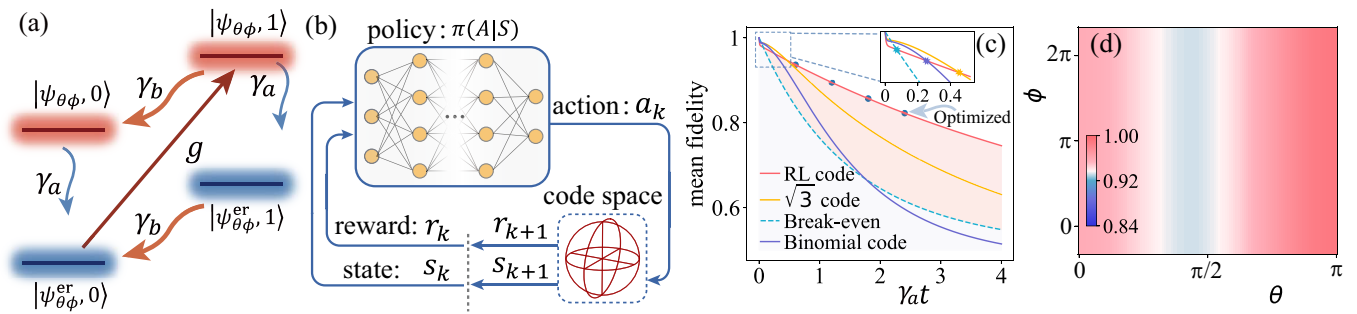}
	\caption{\label{figee}
		(a) Energy level diagram illustrating the approximate AQEC process. {Induced by a single-photon loss,} the encoded state $\vert\psi_{\theta \phi}\rangle$ {undergoes one out of two possible error correction cycles:} $\vert\psi_{\theta \phi},0\rangle\stackrel{\gamma_a}{\longrightarrow}\vert\psi_{\theta \phi}^{\text{er}},0\rangle\stackrel{g}{\longrightarrow}\vert\psi_{\theta \phi},1\rangle\stackrel{\gamma_b}{\longrightarrow}\vert\psi_{\theta \phi},0\rangle$ {or} $\vert\psi_{\theta \phi},0\rangle\stackrel{\gamma_a}{\longrightarrow}\vert\psi_{\theta \phi}^{\text{er}},0\rangle\stackrel{g}{\longrightarrow}\vert\psi_{\theta \phi},1\rangle\stackrel{\gamma_{a}}{\longrightarrow}\vert\psi^{\text{er}}_{\theta \phi},1\rangle\stackrel{\gamma_b}{\longrightarrow}\vert\psi^{\text{er}}_{\theta \phi},0\rangle\stackrel{g}{\longrightarrow}\vert\psi_{\theta \phi},1\rangle\stackrel{\gamma_b}{\longrightarrow}\vert\psi_{\theta \phi},0\rangle$, where the parameters satisfy $\gamma_a, g\ll\gamma_b$  and $\gamma_a\ll g$.
		(b) Diagram: the cyclic learning process between the RL agent and the encoded system. The policy function $\pi(A\vert S)$ selects an action based on the current state and reward, and the action acts on the encoded system to generate the new reward $r_{k+1}$ and state $s_{k+1}$.
		The states and rewards collected during the sampling process are fed back to the policy network for updating the networks' parameters.			
		 (c) {Comparison of} the AQEC performance of: RL code; the lowest-order binomial code; $\sqrt{3}$ code; break-even. {The blue round dots show the optimal mean fidelities obtained from the RL algorithm for different fixed optimization times. The inset highlights an initial transition period where the dynamics is dominated by the single-photon loss.} (d) {State-dependent} fidelity $F(\theta,\phi, t)$ {for the RL code} versus {the} angles $\theta$ and $\phi$ at {$t=0.6/\gamma_{a}$. Other parameters are $\gamma_b/\gamma_a=1750$ and $g/\gamma_a=400$.}} 
	\label{fig1}
\end{figure*} 
{AQEC employs engineered dissipation to recover from the errors that occur due to natural decay processes}.~In brief, engineered Lindblad operators $\sum_k\mathcal{D}[L_{\text{eng},k}]$ are introduced such that the overall dissipative channel $\mathcal{M}[\cdot]$, which also includes the natural Lindblad operators $\sum_i\mathcal{D}[L_{\text{nat},i}]$, minimizes the growth of the infidelity between an arbitrary {qubit} state $\rho_{t_0}(\theta,\phi)=\vert \psi_{\theta \phi}\rangle\langle \psi_{\theta \phi}\vert$ in the code space and the {evolved} state $\rho_t(\theta,\phi)=\mathcal{M}[\rho_{t_0}(\theta,\phi)]$.~This happens by enlarging the effect of $\sum_k\mathcal{D}[L_{\text{eng},k}]$ \cite{Lebreuilly2021Mar}, where $\mathcal{D}[x]=2x\rho x^{\dagger}-x^{\dagger}x\rho-\rho x^{\dagger}x$, and the angles $\theta$ and $\phi$ parametrize an arbitrary qubit state in the Bloch-sphere representation.

The mean fidelity $\bar{F}(\vert 0_{\text{L}}\rangle,\vert 1_{\text{L}}\rangle)$, and hence the performance of the AQEC, {depends on the choice of the logical codewords} $\vert 0_{\text{L}}\rangle$ and $\vert 1_{\text{L}}\rangle$. If the logical {codewords satisfy} the KL condition \cite{PhysRevLett.84.2525},
\begin{align}
	\langle u_{\text{L}}\vert L^{\dagger}_{\text{nat},j}L_{\text{nat},i}\vert v_{\text{L}}\rangle=\alpha_{ji}\delta_{uv},~~u,v\in \lbrace 0,1\rbrace,
	\label{eq1}
\end{align}
where $\alpha_{ji}$ {are the elements} of a Hermitian matrix, one can, {in principle,} achieve exact QEC. However, {in the context of AQEC for bosonic codes, where $L_{\text{nat},i}\in \lbrace I, a\rbrace$ \cite{Lescanne2020May,RevModPhys.93.025005,Gu2017Nov}, the exact implementation of the KL conditions can be challenging. For instance, the necessity to engineer multiple Lindblad operators, as is the case with the binomial and the T4C code, dramatically increases experimental complexity. A similar challenge occurs if multiple high-order nonlinear interactions, along with multiple control fields, are required, as for the $\sqrt{3}$ code.}

It has been shown that approximate QEC (i.e., the codewords partially satisfy the KL condition) can combine experimental feasibility with good QEC performance \cite{Herrera-Marti2015Nov,Kwon2022Apr,Brandao2019Sep,Tzitrin2020Mar}. Following this line of research, we relax the KL condition {by discarding the constraint that} $\langle 1_{\text{L}}\vert L^{\dagger}_{\text{nat},i}L_{\text{nat},i}\vert 1_{\text{L}}\rangle=\langle 0_{\text{L}}\vert L^{\dagger}_{\text{nat},i}L_{\text{nat},i}\vert 0_{\text{L}}\rangle$, i.e., {we allow} the logical code {words to exhibit} different error probabilities. {General codewords (selecting the even subspace) that satisfy the remaining KL conditions can then be parametrized as} 
\begin{equation}
	\vert 0_{\text{L}}\rangle=\sum_{n=0}c_{n}^{(0)}\vert 4n\rangle, ~~~~
	\vert 1_{\text{L}}\rangle=\sum_{n=0} c_{n}^{(1)}\vert 4n+2\rangle,
\label{eq2}
\end{equation}
where the $c^{(0)}_n$ and $c^{(1)}_n$ {denote} undetermined real coefficients and satisfy $\sum_{n}\vert c_{n}^{(u)}\vert^2=1$ ($u=0,1$). {Our goal is now to find coefficients $c^{(0)}_n$ and $c^{(1)}_n$ that optimize the performance of the corresponding AQEC with only a single, fixed} QEC jump operator, 
\begin{equation}
	\begin{aligned}
	L_{\text{eng}}&=L_{\text{o}}\left\lbrace\operatorname{Tr}\left[L^{\dagger}_{\text{o}}L_{\text{o}}\right]\right\rbrace^{-1/2}, 	 \\
    L_{\text{o}}&=\vert 0_{\text{L}}\rangle\langle 0_{\text{er}}\vert+\vert 1_{\text{L}}\rangle\langle 1_{\text{er}}\vert,	
\end{aligned}
\label{eq3}
\end{equation}
where $\vert u_{\text{er}}\rangle=a\vert u_{\text{L}}\rangle/\xi_u$ $(u=0,1)$ denotes the basis of the error space and $\xi_u=\sqrt{\langle u_{\text{L}}\vert a^{\dagger}a\vert u_{\text{L}}\rangle}$ is the normalization coefficient. The QEC jump operator $L_{\text{eng}}$ steers the encoded quantum state from the error space {back} into the code space and can be expanded as
 \begin{equation}
 L_{\text{eng}}=\sum_{\vert d_l\vert\leq d}\sum_n\lambda_{nd_l}\vert n \rangle\langle n+d_l\vert,
 \label{eq4}
 \end{equation}
where $d$ is the Hamiltonian distance.  
 
{We can model the engineered dissipation through a system-environment interaction by introducing a lossy auxiliary qubit and the coupling} $H_{\text{eff}}=g(L_{\text{eng}}\sigma_++L^{\dagger}_{\text{eng}}\sigma_-)$, {which results in the} engineered Lindblad superoperator $\mathcal{D}[L_{\text{eng}}]$ {when the qubit is traced out}. The evolution of the whole system is {then governed} by the master equation ($\hbar=1$) 
\begin{equation}
\frac{d\rho}{dt}=-i[H_{\text{eff}}, \rho]+\frac{\gamma_a}{2}\mathcal{D}[a]+\frac{\gamma_b}{2}\mathcal{D}[\sigma_-],
\label{eq5}
\end{equation}
where $\gamma_a$ {denotes} the single-photon loss {rate} of mode $a$ and $\gamma_b$ {denotes} the decay rate of the auxiliary qubit. {We assume that} the parameters satisfy $\gamma_a, g\ll\gamma_b$ and $\gamma_a\ll g$, {which promotes the unidirectional transition from the error space to the code space effected by $L^{\dagger}_{\text{eng}}$}. As shown in Fig.~\ref{fig1}(a), the {recovery} process is {then} as follows: $\vert\psi_{\theta \phi},0\rangle\stackrel{\gamma_a}{\longrightarrow}\vert\psi_{\theta \phi}^{\text{er}},0\rangle\stackrel{g}{\longrightarrow}\vert\psi_{\theta \phi},1\rangle\stackrel{\gamma_b}{\longrightarrow}\vert\psi_{\theta \phi},0\rangle$ and $\vert\psi_{\theta \phi},0\rangle\stackrel{\gamma_a}{\longrightarrow}\vert\psi_{\theta \phi}^{\text{er}},0\rangle\stackrel{g}{\longrightarrow}\vert\psi_{\theta \phi},1\rangle\stackrel{\gamma_{a}}{\longrightarrow}\vert\psi^{\text{er}}_{\theta \phi},1\rangle\stackrel{\gamma_b}{\longrightarrow}\vert\psi^{\text{er}}_{\theta \phi},0\rangle\stackrel{g}{\longrightarrow}\vert\psi_{\theta \phi},1\rangle\stackrel{\gamma_b}{\longrightarrow}\vert\psi_{\theta \phi},0\rangle$. {That is, when an error occurs, the system returns to the logical state $\vert\psi_{\theta\phi}, 0\rangle$ by transitioning through one out of two possible error correction cycles.}

{{\it Optimal code space.}---Finding the optimal coefficients $c^{(0)}_n$ and $c^{(1)}_n$, such that the codewords (\ref{eq2}) maximize the mean fidelity $\bar{F}(\vert 0_{\text{L}}\rangle,\vert 1_{\text{L}}\rangle)$, at some fixed reference time, represents a complex optimization problem that we solve using reinforcement learning (RL). In brief, each episode is divided} into a finite number of steps $k=1, 2, \cdots, K$. As shown in Fig.~\ref{fig1}(b), at each step $k$, the agent observes the current state $s_k\in S$, and chooses an action $a_k \in A$ according to the policy $\pi(A\vert S)$. {The action corresponds to a coefficient vector [$c^{(0)}_n$, $c^{(1)}_n$], and the state is described by the  fidelity $F(\theta,\phi,t)=\operatorname{Tr}[\rho_{t_0}(\theta,\phi)\rho_t(\theta,\phi)]$, with $\theta\in\{0,\pi/2,\pi\}$ and $\phi\in\{0,\pi/2,\pi,3\pi/2\}$. After each action, the agent receives a reward $r_{k+1}$ and arrives at a new state $s_{k+1}$, where the reward $r_k$ is set to maximize the difference between the mean fidelity of the logical space and the break-even point. State and reward are obtained by simulating the master equation~(\ref{eq5}) until $\gamma_at=0.6$ with QuTiP \cite{Johansson2013Apr, Johansson2012Aug}, where the parameters $g/\gamma_a=400$, $\gamma_b/\gamma_a=1750$ are chosen to maximize the cooperativity $C=g^2/\gamma_a\gamma_b \approx 91.4$, while remaining experimentally realistic.} The states and rewards collected during the sampling process are fed back to the proximal policy optimization algorithm to update the policy $\pi(A\vert S)$ \cite{Schulman2017Jul}, which is achieved by Ray \cite{BibEntry2022Jun} (See Supplementary Materials for the more information~\cite{Supplement}).

The optimal code {basis} found by the RL algorithm, {delivering a mean fidelity $\approx 0.95$ at $\gamma_a t = 0.6$ and thus significantly exceeding the break-even point at $\approx 0.84$}, {consists of the} Fock {states}: {$|0_{\rm L}\rangle\approx|4\rangle$ and $|1_{\rm L}\rangle\approx|2\rangle$}, i.e., $c^{(0)}_1\approx 1$, $c^{(1)}_0\approx1$, $c^{(0)}_0\approx c^{(1)}_1\approx0$ [Note that we {truncated} the code space {at} 6 photons. {Moreover, we stress that different optimization times $\gamma_a t$ yield consistent outcomes, cf.~Fig.~\ref{fig1}(c)}].~These simple code words can be conveniently prepared in existing experimental setups, e.g., superconducting quantum systems \cite{YuxiLiu_2004,Hofheinz2009May,Hofheinz2008Jul,Supplement11,Phys.Rev.Lett.76.3108}.~Moreover, the {respective} QEC jump operator $L_{\text{eng}}\propto \vert 2\rangle\langle 1\vert+\vert 4\rangle\langle 3\vert$ has {the shortest possible} Hamiltonian distance $d=1$, {implying that} the QEC Hamiltonian can be {efficiently} implemented without {the need for multiple} nonlinear interactions and control fields \cite{PRXQuantum.3.020302, Gertler2021Feb}.~The {Hamiltonian} distance of the codewords is 2, {resulting in} single-qubit gates with $d_g=2$. For {instance}, the logical Pauli operators {are implemented as} $\sigma_{x}=\vert 2\rangle\langle 4\vert+\vert 4\rangle\langle 2\vert$, {$\sigma_{y}=i(\vert 4\rangle\langle 2\vert-\vert 2\rangle\langle 4\vert)$, and $\sigma_{z}=\vert 2\rangle\langle 2\vert-\vert 4\rangle\langle 4\vert$.}
{For comparison,} the lowest-order binomial code {requires} $d_g=4$ and the $\sqrt{3}$ code $d_g=6$. {Consequently}, single-qubit gates {in the RL code space} only require second-order nonlinearity, instead of the fourth-order or sixth-order {nonlinearities of} previous proposals.

In Fig.~\ref{fig1}(c), we compare the mean fidelities of the RL code, $\sqrt{3}$ code, and the lowest-order binomial code, where each code is complemented by a single QEC jump operator Eq.~(\ref{eq3}). 
We find that the RL code surpasses {break-even} and {both} other codes, {with} the {performance} advantage growing in time. The mean fidelity of the RL code exceeds the break-even point by $\approx 36\%$ at $\gamma_{a}t=4$.

Note that there is a short transition period where the RL code performs poorer than the other codes. This is because initially, when the system lives entirely in the code space, the jump operator Eq.~(\ref{eq3}) is not yet effective and the dynamics is entirely driven by the detrimental single-photon loss~\cite{New.J.Phys.15.035014}. The associated decay rate is proportional to the mean photon number, which is highest for the RL code in our comparison ($\bar{n}=3$ for the RL code, compared to $\bar{n}=2$ for the binomial code, $\bar{n}=\sqrt{3}$ for the $\sqrt{3}$ code, and $\bar{n}=0.5$ for the break-even code). Only after this transition period the jump operator Eq.~(\ref{eq3}) becomes effective and the RL code can unfold its performance advantage over the other codes. Note that the dip at the onset of the transition period reflects the reversible fraction of the fidelity loss and can be removed by using a quantum trajectory-resolved and temporally coarse-grained fidelity $F^*_{\tau}(t)$~\cite{New.J.Phys.15.035014, Supplement}.

To demonstrate that the AQEC can {efficiently} protect all quantum states {in} the RL code space, we {evaluate in Fig.~\ref{fig1}(d) the fidelity} $F(\theta,\phi,t)$ at {$t = 0.6/\gamma_{a} $}, where $\theta$ and $\phi$ are the Bloch-sphere angles. {We find} that the fidelity {remains for all states} within the range $[0.93,1]$, {well} above the break-even point $0.84$.

{\it {Protection of the logical qubit}}---To better understand {the level of protection provided by our AQEC scheme}, we {solve the master equation in} the limit $\gamma_a,g\ll\gamma_{b}$ and $g^2/(\gamma_{a}\gamma_b)\gg1$ ({a detailed derivation can be found} in the Supplemental Material~\cite{Supplement}). {The approximate solution, after tracing out the auxiliary qubit and expressed in terms of the first five Fock states, reads} 
\begin{equation}
	\rho_a(t)\approx\left(\begin{array}{ccccc}
		0 & 0 & 0 & 0 & 0\\
		0 & 0 & 0 & 0 & 0\\
		0 & 0 & \rho_{22}(0) & 0 & \rho_{24}(0)e^{-u\gamma_{a}t}\\
		0 & 0 & 0 & 0 & 0\\
		0 & 0 & \rho_{42}(0)e^{-u\gamma_{a}t} & 0 & \rho_{44}(0)
	\end{array}\right),  
	\label{seq11} 
\end{equation}
{with the initial state elements $\rho_{ij}(0)$. The effective protection factor $u=3-2\sqrt{2}\approx 0.17$ captures the reduction of the residual dephasing rate compared to the $|0\rangle$ and $|1\rangle$ encoding. The resulting mean fidelity is upper bounded by} 
\begin{eqnarray}
	\bar{F}(t)=\frac{2}{3}+\frac{1}{3}\exp(- u\gamma_a t), ~~ u=3-2\sqrt{2}\approx 0.17,
	\label{eq7}
\end{eqnarray}
{and thus clearly outperforms the mean fidelity of the break-even} 
\begin{eqnarray}
\bar{F}_{\text{be}}(t)=\frac{1}{6}\left[\exp{(-\gamma_a t)}+2\exp{(-\frac{\gamma_a t}{2})}+3\right].
\end{eqnarray}
We infer from Eq.~(\ref{eq7}) that, if the {cooperativity $C$} is large enough, the performance of AQEC depends only on the {evaluation time} $\gamma_at$. If we choose $\gamma_at= 0.17$ ({e.g.}, $\gamma_a = 0.1$~kHz, $t = 1.7$~ms), the mean fidelity {is} $99.05\%$, which significantly surpasses the break-even point ($94.68\%$). The infidelity reduction of {the} RL code compared to {the} break-even is $[1-\bar{F}(t)]/[1-\bar{F}_{\text{be}}(t)]\approx 0.17$, {corresponding to the} gain $G=1/u\approx5.83$ (near to 6 times the T4C code \cite{Gertler2021Feb}).~Therefore, {the approximate AQEC with} the RL code significantly surpasses {the} break-even point with an infidelity reduction of over 80\%. {We clarify that, similar to other AQEC schemes, leakage out of the error space ultimately limits the protection time $\gamma_at$, and eventually the system will end up in the ground state.} 

 \begin{figure}[bt]
	\includegraphics[width=3.1in]{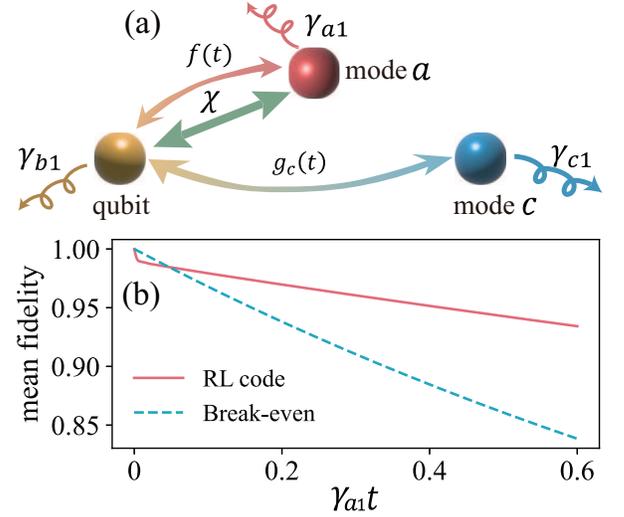}
	\caption{(a) {Proposed system-environment coupling.~Blue sideband transitions between a qubit and the encoding mode $a$ are selected by the control field $f(t)$. Red sideband transitions, mediated by the control field $g_c(t)$, transfer the excitation of the qubit to the high-decay mode $c$, where it dissipates into the environment.} (b) Mean fidelity of {the} RL code versus time for {this coupling} model under parameters $\omega_a/2\pi=3.5$~GHz, $\omega_b/2\pi=\omega_c/2\pi=5$~GHz, $\chi/2\pi=3$~MHz, $\alpha_0/2\pi=0.05$~MHz, $\alpha_1/2\pi=0.07$~MHz, $\gamma_{a1}/2\pi=0.2$~kHz, $\gamma_{b1}/2\pi=2$~kHz, and $\gamma_{c1}/2\pi=0.12$~MHz.} 
	\label{fig2}
\end{figure}
{\it {Coupling engineering.}}---
{In contrast to previous AQEC proposals \cite{PRXQuantum.3.020302, Gertler2021Feb}, where complex codewords necessitate multiple nonlinear interactions and complicated control fields, AQEC based on the RL code only requires a Hamiltonian distance $d=1$, which can be realized with a comparatively simple setting.} {Figure~\ref{fig2}(a) depicts a possible scheme, where} the encoding mode $a$ is intermediately coupled to a dissipative mode $c$ through a qubit. The {corresponding system} Hamiltonian {reads}
\begin{equation}
	\begin{split}
	H=&\omega_aa^{\dagger}a+\frac{\omega_b}{2}\sigma_{z}+\omega_cc^{\dagger}c+f(t)(a+a^{\dagger})\sigma_{x}
	\\
	&+g_c(t)(c^{\dagger}+c)\sigma_{x}+\frac{\chi}{2}a^{\dagger}a\sigma_{z},
	\end{split}
	\label{seq22}
\end{equation}
and the control fields are
\begin{equation}
    \begin{aligned}
	    &f(t) \!=\!\frac{2\alpha_0}{\sqrt{2}}\cos\left[(\omega_s+\frac{3\chi }{2})t\right]+\frac{2\alpha_0}{\sqrt{4}}\cos\left[(\omega_s+\frac{7\chi}{2})t\right], \\
	    &g_c(t)\!=\!2\alpha_1\cos(2\chi t)+2\alpha_1\cos(4\chi t),
    \end{aligned}
    \label{eq12}
\end{equation}
where $\omega_{i}$ ($i=a,b,c$) are the resonant frequencies of the modes and {the} qubit ($\omega_b=\omega_c$), $\omega_s=\omega_a+\omega_b$ is the resonant frequency for the blue sideband transitions, $\chi$ is the nonlinear coefficient, and $\alpha_j$ ($j=0,1$)  are the {strengths} of the control fields ({For detailed} information about the control fields, see the supplemental materials~\cite{Supplement}). We can {recover} the master equation(\ref{eq5}) by {adiabatically eliminating} the high-decay mode $c$ under the {conditions} $\omega_a,\omega_b,\omega_c\gg\chi$, {and} $\gamma_{c1}\gg\alpha_1\geq\alpha_0\gg\gamma_{a1}, \gamma_{b1}$, where $\gamma_{i1}$ ($i=a,b,c$) are the {respective} decay rates. The controllable coupling between {qubits} and bosonic modes has been {thoroughly investigated both in theory and in experiment} \cite{PhysRevLett.99.050501, PhysRevLett.119.150502, Roushan2017Feb, Bresque2021Mar, Campagne-Ibarcq2018May,Niemczyk2010Oct, Leek2009May, PhysRevB.76.144518, Niemczyk2010Oct}.~The nonlinear term $a^{\dagger}a\sigma_{z}$ can be effectively obtained by a linear coupling between mode and qubit, $\beta(a^{\dagger}+a)\sigma_x$, with a large detuning $\Delta=\vert\omega_a-\omega\vert\gg\beta$, where the dispersive coefficient $\chi=g^2/\Delta$ can be larger than $7\times 2\pi$~MHz  \cite{Schuster2007Feb,Majer2007Sep,PhysRevB.68.064509,Cai2021Feb,Xiang2013Apr,PhysRevA.81.042311,PhysRevA.80.033846,PhysRevLett.115.180404,Brown2022Jul}. {It is thus possible to realize the scheme with linearly coupled systems.} {A} low single-photon loss {of} $0.118\times 2\pi$ kHz has been achieved in {experiments with 3D coaxial cavities}, and the decay of {the transmon qubits can be suppressed to} $1.8\times 2\pi$ kHz \cite{Heeres2017Jul,Rosenblum2018Jul,Wang2016May}. We simulate the mean fidelity of the full quantum system in Fig.~\ref{fig2}(b) with typical experimental parameters in superconducting circuits, $\omega_a/2\pi=3.5$~GHz, $\omega_b/2\pi=\omega_{c1}/2\pi=5$~GHz, $\gamma_{c1}/2\pi=0.12$~MHz, $\chi/2\pi=3$~MHz, $\alpha_0/2\pi=0.05$~MHz, $\alpha_1/2\pi=0.07$~MHz, $\gamma_{a1}/2\pi=0.2$~kHz, and $\gamma_b/2\pi=2$~kHz.~We find that the mean fidelity of the RL code {by} far surpasses break-even threshold, and the model {may be realized} by engineering the coupling between a 3D coaxial cavity, a transmon qubit, and a dissipative cavity.

{\it Discussion and conclusion}---We propose a bosonic code space for approximate AQEC with the shortest {possible} Hamiltonian distance $d=1$, {allowing for significantly reduced model complexity. Nevertheless, our scheme comfortably surpasses break-even threshold, outperforming other AQEC schemes that rely on larger Hamiltonian distances and disproving a previous claim that surpassing break-even threshold requires $d\geq2$}. The codewords consist of Fock states rather than complex superposition states, and single-qubit logic gates in the RL code space {are implemented with a} maximum distance of $d_g=2$. The mean fidelity of {the} RL code can exceed break-even point by more than $36\%$ at $\gamma_at=4$, and the {expected} gain $G=1/(3-2\sqrt{2})\approx5.83$ is {more than twice the currently best experimental value} $G= 2.27 \pm 0.07$ \cite{Sivak2022Nov}.
{This demonstrates that, despite our AQEC scheme being approximate, which upper bounds the level of protection, it delivers logical qubits with significantly improved quality and thus may greatly facilitate subsequent QEC steps toward fully fault-tolerant qubits.}

~\\ 

We would like to acknowledge the help of Professor~Jie-Qiao Liao and the valuable suggestions of Dr.~Wei Qin, Dr.~Ye-Hong Chen, Dr.~Fabrizio Minganti, and Dr.~Ran Huang. F.N. is supported in part by 
Nippon Telegraph and Telephone Corporation (NTT) Research, 
the Japan Science and Technology Agency (JST) 
[via the Quantum Leap Flagship Program (Q-LEAP) and the Moonshot R\&D Grant No. JPMJMS2061],
the Asian Office of Aerospace Research and Development (AOARD) (via Grant No. FA2386-20-1-4069), and 
the Foundational Questions Institute Fund (FQXi) via Grant No. FQXi-IAF19-06.

\bibliographystyle{apsrev4-1}
%







\end{document}


\title{Supplementary Materials for ``Approximate Autonomous Quantum Error Correction with Reinforcement Learning''}
\author{Yexiong Zeng\orcid{0000-0001-9165-3995}}
\affiliation{Theoretical Quantum Physics Laboratory, Cluster for Pioneering Research, RIKEN, Wakoshi, Saitama 351-0198, Japan}
\affiliation{Quantum Computing Center, RIKEN, Wakoshi, Saitama 351-0198, Japan}

\author{Zheng-Yang Zhou}%
\affiliation{Theoretical Quantum Physics Laboratory, Cluster for Pioneering Research, RIKEN, Wakoshi, Saitama 351-0198, Japan}%

\author{Enrico Rinaldi\orcid{0000-0003-4134-809X}}
\affiliation{Quantinuum K.K., Otemachi Financial City Grand Cube 3F, 1-9-2 Otemachi, Chiyoda-ku, Tokyo, Japan}
\affiliation{Theoretical Quantum Physics Laboratory, Cluster for Pioneering Research, RIKEN, Wakoshi, Saitama 351-0198, Japan}
\affiliation{Quantum Computing Center, RIKEN, Wakoshi, Saitama 351-0198, Japan}
\affiliation{Department of Physics, University of Michigan, Ann Arbor, Michigan, 48109-1040, USA}
\affiliation{Interdisciplinary Theoretical and Mathematical Sciences Program (iTHEMS), RIKEN, Wakoshi, Saitama 351-0198,
	Japan}

\author{Clemens Gneiting\orcid{0000-0001-9686-9277}}
\altaffiliation[clemens.gneiting@riken.jp]{}
\affiliation{Theoretical Quantum Physics Laboratory, Cluster for Pioneering Research, RIKEN, Wakoshi, Saitama 351-0198, Japan}
\affiliation{Quantum Computing Center, RIKEN, Wakoshi, Saitama 351-0198, Japan}

\author{Franco Nori\orcid{0000-0003-3682-7432}}
\altaffiliation[fnori@riken.jp]{}
\affiliation{Theoretical Quantum Physics Laboratory, Cluster for Pioneering Research, RIKEN, Wakoshi, Saitama 351-0198, Japan}
\affiliation{Quantum Computing Center, RIKEN, Wakoshi, Saitama 351-0198, Japan}
\affiliation{Department of Physics, University of Michigan, Ann Arbor, Michigan, 48109-1040, USA}


\maketitle
\tableofcontents
\title{Approximate Autonomous Quantum Error Correction with Reinforcement Learning}%
\section{Mean fidelity of the code space}
In this section, we determine the fidelity between initial and evolved code states averaged over the code space. An arbitrary state in the code space (i.e., the initial state) can be expanded as   
\begin{equation}
	\left|\psi_{\theta \phi}\right\rangle=\cos \frac{\theta}{2}\left|0_{\text{L}}\right\rangle+e^{i \phi} \sin \frac{\theta}{2}\left|1_{\text{L}}\right\rangle,
	\label{seq1}
\end{equation}
where $\vert 0_{\text{L}}\rangle$, $\vert 1_{\text{L}}\rangle$ are the logical codewords, and $\theta$, $\phi$ are the familiar Bloch sphere angles.
A linear, positive definite, and trace-preserving map $\mathcal{M}[\cdot]$  evolves the initial state $\rho_{t_0}(\theta,\phi)=\left|\psi_{\theta \phi}\rangle\langle \psi_{\theta \phi}\right\vert$ to the quantum state $\rho_t(\theta,\phi)=\mathcal{M}[\rho_{t_0}(\theta,\phi)]$ at time $t$. With the fidelity between the evolved state $\rho_t(\theta,\phi)$ and the initial state $\rho_{t_0}(\theta,\phi)$ given by $F(\theta,\phi,t)=\operatorname{Tr}[\rho_{t_0}(\theta,\phi)\rho_t(\theta,\phi)]$, the mean fidelity of the code space can be written as 
\begin{equation}
	\bar{F}(t)=\frac{1}{4\pi}\int_{\Omega}	F(\theta,\phi,t)~\text{d}\Omega.
	\label{seq2}
\end{equation}
Next, we expand the initial state in terms of the logical Pauli operators 
\begin{equation}
	\begin{aligned}
		\rho_{t_0}(\theta,\phi)=\left|\psi_{\theta, \phi}\right\rangle\left\langle\psi_{\theta, \phi}\right| =\sum_{j=0, x, y, z} c_{j}(\theta, \phi) \frac{\sigma_{j}}{2},
	\end{aligned}
	\label{seq3}
\end{equation}
where
\begin{equation}
  \begin{aligned}
  	~~~~~&	\sigma_{0}=\vert 0_{\text{L}}\rangle\langle 0_{\text{L}}\vert+\vert 1_{\text{L}}\rangle\langle 1_{\text{L}}\vert,~~~~~
  	\sigma_{x}=\vert 0_{\text{L}}\rangle\langle 1_{\text{L}}\vert+\vert 1_{\text{L}}\rangle\langle 0_{\text{L}}\vert, 
  	\\
  	~~~~~&	\sigma_{y}=i(\vert 0_{\text{L}}\rangle\langle 1_{\text{L}}\vert-\vert 1_{\text{L}}\rangle\langle 0_{\text{L}}\vert), 	
  	~~
  	\sigma_{z}=\vert 1_{\text{L}}\rangle\langle 1_{\text{L}}\vert-\vert 0_{\text{L}}\rangle\langle 0_{\text{L}}\vert,	
  \end{aligned}
\label{seq4}
\end{equation}
and the elements of the coherent Bloch vector are
\begin{equation}
	\begin{aligned}
	c_0(\theta,\phi)=1, ~~c_x(\theta,\phi)=\sin\theta\cos\phi, ~~
	c_y(\theta,\phi)=\sin\theta\sin\phi,~~ c_z(\theta,\phi)=\cos\theta.
	\end{aligned}
\label{seq5}
\end{equation}
By combining Eqs.~(\ref{seq2}), (\ref{seq3}), (\ref{seq4}), (\ref{seq5}),  we can reformulate the mean fidelity as
\begin{equation}
	\begin{aligned}
		\bar{F}(t) &=\frac{1}{4 \pi} \int_{0}^{\pi} \int_{0}^{2 \pi} \operatorname{Tr}\left(\sum_{j} c_{j}(\theta, \phi) \frac{\sigma_{j}}{2}  \mathcal{M}\left[\sum_{k} c_{k}(\theta, \phi) \frac{\sigma_{k}}{2}\right]\right) \sin \theta~ \mathrm{d} \phi \mathrm{d} \theta \\
		&=\sum_{j k}\frac{1}{4 \pi} \int_{\theta} \int_{\phi} c_{j} c_{k} \sin \theta ~\mathrm{d} \phi \mathrm{d} \theta \operatorname{Tr}\left( \frac{\sigma_{j}}{2} \mathcal{M}\left[\frac{\sigma_{k}}{2}\right]\right)\\
		&=\frac{1}{6} \sum_{j=\pm x, \pm y, \pm z}\operatorname{Tr}\left( \rho_{j}  \mathcal{M}\left[\rho_{j}\right]\right) ,
	\end{aligned}
	\label{seq6}
\end{equation}
where we have defined $\rho_0=\sigma_0/2$ and $\rho_{\pm j}=(\sigma_0 \pm \sigma_{j})/2$ ($j=x,y,z$). This result is consistent with previous work \cite{Wang2022Apr}.

\section{Search for the optimal code space with reinforcement learning }
\begin{figure}[bt]
	\includegraphics[width=5 in]{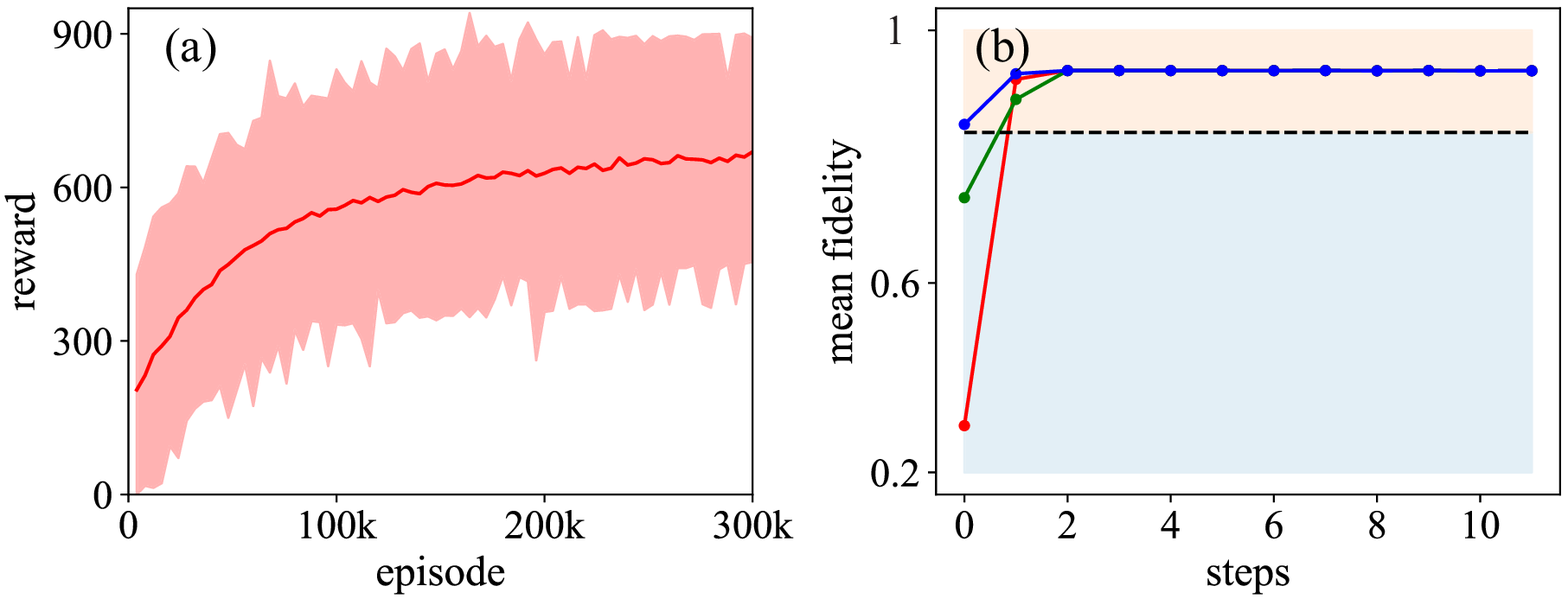}
	\caption{(a) Plot of the maximum (upper bound of the shaded area),  average (solid red line), and minimum rewards (lower bound of the shaded area)  as functions of the episodes during the training process, where the dynamic evolution time is {$\gamma_{a}t=0.6$}. (b) Evolution of the mean fidelity with the number of steps for three randomly selected episodes. Other parameters are $g/\gamma_a=400$ and $\gamma_b/\gamma_a=1750$ (i.e., $C\approx91.4$).} 
	\label{sfig1q}
\end{figure}
Finding the optimal coefficients $c^{(0)}_n$ and $c^{(1)}_n$, such that the codewords, Eq.~(2) of the main text, maximize the mean fidelity $\bar{F}(\vert 0_{\text{L}}\rangle,\vert 1_{\text{L}}\rangle)$ at some fixed reference time, represents a complex optimization problem that we solve using reinforcement learning (RL). In brief, we divide each episode into a finite number of steps $k=1, 2, \cdots, K$. At step $k$ the agent observes the state $s_k\in S$ of the surrounding environment and chooses an action $a_k$ according to the policy $\pi(A\vert S)$. Then the agent obtains the new state $s_{k+1}$, and the environment returns a reward $r_{k+1}$. The policy $\pi(A\vert S)$ is updated via the experience data to maximize the accumulated reward $R$. The state $s_k$ contains six observations $\operatorname{Tr}(\rho_j \mathcal{M}(\rho_{j}))$, $j=\pm x, \pm y, \pm z$ [see Eq.(\ref{seq6})] that can be calculated by using the toolkit Qutip \cite{Johansson2012Aug,Johansson2013Apr}. The action of the agent consists of the vector of coefficients $[c^{(0)}_n, c^{(1)}_n]$. The reward $r_{k+1}$ is proportional to the difference between the mean fidelities of the code space obtained by the policy function and of the break-even point,  $\epsilon_k=\bar{F}_{k}(\vert 0_L\rangle,\vert 1_L\rangle)-\bar{F}(\vert 0\rangle,\vert 1\rangle)$. The specific reward scheme is as follows: if $\epsilon_{k+1}>0$ and $\epsilon_{k+1}>\epsilon_{k}$, the reward is $1000\epsilon_{k+1}$; if $\epsilon_{k+1}>0$ and $\epsilon_{k+1}<\epsilon_{k}$, the reward is $100\epsilon_{k+1}$; if $\epsilon_{k}\leq 0$, the reward is $0$. 

We truncate the code space to 6 photons and set the maximum number of steps per episode to $K=11$. The parameters for the simulation of the dynamics are chosen as $g/\gamma_a=400$, $\gamma_b/\gamma_a=1750$ (i.e., $C\approx91.4$), $\gamma_{a}=0.02$~MHz, and $\gamma_at=0.6$, in agreement with current experimental conditions. We apply the proximal policy optimization algorithm to optimize the policy~\cite{Schulman2017Jul}, which is achieved by using the Python toolkit Ray \cite{BibEntry2022Jun}. The parameters of the neural networks are the default ones offered by Ray. 

As shown in Fig.~\ref{sfig1q}(a), the mean reward approximately converges to a constant after about 200k episodes. Meanwhile, the max reward is obtained when the coefficient is $c^{(0)}_1\approx 1$, $c^{(1)}_0\approx1$, and $c^{(0)}_0\approx c^{(1)}_1\approx0$. We thus conclude that the optimal code space consists of the Fock states $\vert 2\rangle$ and $\vert 4\rangle$.

To demonstrate that different initial states converge to the same mean fidelity, we display in Fig.~\ref{sfig1q}(b) three random episodes. One can easily see that all three episodes converge towards the same mean fidelity (well above the break-even point of 0.84), in each case yielding the same optimal codewords, the Fock states $\vert 2\rangle$ and $\vert 4\rangle$, as described by the RL code.

\section{Assessing the performance of the approximate AQEC}

Here we analytically assess the performance of the approximate AQEC. First, we solve the dynamical evolution for an arbitrary initial code state, and then calculate the mean fidelity. The dynamic evolution of the hybrid system composed of the encoding and the auxiliary mode is governed by the master equation
\begin{equation}
	\frac{d\rho}{dt}=-i[H_I, \rho]+\frac{\gamma_a}{2}\mathcal{D}[a]+\frac{\gamma_b}{2}\mathcal{D}[\sigma_-],
	\label{seq7}
\end{equation}
where the Hamiltonian of the hybrid system is $H_I=g(L_{\text{eng}}\sigma_{+}+L^{\dagger}_{\text{eng}}\sigma_-)$, and the resulting error correction operator is the same as in the main text.

If the parameters satisfy the conditions $g,\gamma_a\ll\gamma_b$, we can approximately write the density operator of the hybrid system as $\rho(t)=\rho_a(t)\otimes\vert 0\rangle\langle 0\vert$, where $\rho_a(t)$ is the state of the encoding mode and $\vert 0\rangle\langle 0\vert$ is the ground state of the auxiliary qubit.
Therefore, the dynamical evolution of the encoding mode is governed by the effective master equation \cite{Zanardi2016Feb}
\begin{eqnarray}
	\frac{d\rho_a}{dt}=\frac{\gamma_a}{2}\mathcal{D}[a]+\frac{\gamma_a\lambda}{2}\mathcal{D}[L_{\text{eng}}], ~~\lambda=\frac{8\vert g\vert^2}{\gamma_b\gamma_a}=8C.
	\label{seq8}
\end{eqnarray}
Without loss of generality, we can restrict the effective master equation to the first five Fock states (from the ground state to four photons). The density matrix elements then satisfy the equations
\begin{equation}
	\begin{aligned}
		&\frac{1}{\gamma_a}\frac{d\rho_{22}}{dt}=\frac{1}{2}(6\rho_{33}-4\rho_{22}+\lambda\rho_{11}), ~~~~~~~~~ \frac{1}{\gamma_a}\frac{d\rho_{24}}{dt}=\frac{1}{2}(-6\rho_{24}+\lambda\rho_{13}), \nonumber\\
		&\frac{1}{\gamma_a}\frac{d\rho_{44}}{dt}=\frac{1}{2}(-8\rho_{44}+\lambda\rho_{33}),~~~~~~~~~~~~~~~~	\frac{1}{\gamma_a}\frac{d\rho_{42}}{dt}=\frac{1}{2}(-6\rho_{42}+\lambda\rho_{31}),
		\nonumber\\
		&\frac{1}{\gamma_a}\frac{d\rho_{11}}{dt}=\frac{1}{2}(4\rho_{22}-2\rho_{11}-\lambda\text{\ensuremath{\rho_{11}}}), ~~~~~~~~~
		\frac{1}{\gamma_a}\frac{d\rho_{13}}{dt}=\frac{1}{2}(2\sqrt{8}\rho_{24}-4\rho_{13}-\lambda\text{\ensuremath{\rho_{13}}}),
		\nonumber\\
		&\frac{1}{\gamma_a}\frac{d\rho_{33}}{dt}=\frac{1}{2}(8\rho_{44}-6\rho_{33}-\lambda\text{\ensuremath{\rho_{33}}}), ~~~~~~~~~	\frac{1}{\gamma_a}\frac{d\rho_{31}}{dt}=\frac{1}{2}(2\sqrt{8}\rho_{42}-4\rho_{31}-\lambda\text{\ensuremath{\rho_{31}}}),
	\label{seq9}
\end{aligned}
\end{equation}
where we have defined the density matrix as $\rho_a(t)=\sum_{ij}\rho_{ij}(t)\vert i\rangle\langle j\vert$. If we assume $\lambda\gg 1$ and $\lambda\gg \gamma_a t$, we obtain the approximate solution
\begingroup
\allowdisplaybreaks
\begin{align}
		\rho_{00}\approx&  1-\left(\frac{6}{5}+\frac{48}{25\lambda}\right)\rho_{44}(0)\left\lbrace\exp\left[-\frac{4}{\lambda}\gamma_{a}t+O\left(\frac{\gamma_{a}t}{\lambda}\right)^2\right]-\exp\left[-\frac{24}{\lambda}\gamma_{a}t+O\left(\frac{\gamma_{a}t}{\lambda}\right)^2\right]\right\rbrace\nonumber\\
		&-\rho_{22}(0)\exp\left[-\frac{4}{\lambda}\gamma_{a}t+O\left(\frac{\gamma_{a}t}{\lambda}\right)^2\right]-\rho_{44}(0)\exp\left[-\frac{24}{\lambda}\gamma_{a}t+O\left(\frac{\gamma_{a}t}{\lambda}\right)^2\right] + O\left(\frac{1}{\lambda}\right)^2,
		 \nonumber\\
		\rho_{11}\approx&\frac{24}{5 \lambda}\rho_{44}(0)\left\lbrace\exp\left[-\frac{4}{\lambda}\gamma_a t+O\left(\frac{\gamma_{a}t}{\lambda}\right)^2\right]-\exp\left[-\frac{24}{\lambda}\gamma_a t+O\left(\frac{\gamma_{a}t}{\lambda}\right)^2\right]\right\rbrace\nonumber \\
		&+\frac{4}{\lambda}\rho_{22}(0)\exp\left[-\frac{4}{\lambda}\gamma_a t+O\left(\frac{\gamma_at}{\lambda}\right)^2\right]+O\left(\frac{1}{\lambda}\right)^2,
		\nonumber\\
		\rho_{22}\approx&\left(\frac{6}{5}-\frac{72}{25 \lambda}\right)\rho_{44}(0)\left\lbrace\exp\left[-\frac{4}{\lambda}\gamma_a t+O\left(\frac{\gamma_at}{\lambda}\right)^2\right]-\exp\left[-\frac{24}{\lambda}\gamma_a t+O\left(\frac{\gamma_at}{\lambda}\right)^2\right]\right\rbrace\nonumber \\
		&+\left(1-\frac{4}{\lambda}\right)\rho_{22}(0)\exp\left[-\frac{4}{\lambda}\gamma_a t+O\left(\frac{\gamma_at}{\lambda}\right)^2\right]+O\left(\frac{1}{\lambda}\right)^2,
		\nonumber \\
		\rho_{33}\approx& \frac{8}{\lambda}\rho_{44}(0)\exp{\left[-\frac{24}{\lambda}\gamma_at+O\left(\frac{\gamma_at}{\lambda}\right)^2\right]}+O\left(\frac{1}{\lambda}\right)^2,\nonumber
		\\
		\rho_{44}\approx &\left(1-\frac{8}{\lambda}\right)\rho_{44}(0)\exp{\left[-\frac{24}{\lambda}\gamma_at+O\left(\frac{\gamma_at}{\lambda}\right)^2\right]}+O\left(\frac{1}{\lambda}\right)^2,\nonumber
\\		
		\rho_{24}\approx  &\left(1-\frac{4 \sqrt{2}}{\lambda}\right)\rho_{24}(0)\exp\left[\frac{4 (\sqrt{2}-4)\gamma_at}{\lambda}+(2\sqrt{2}-3)\gamma_at+O\left(\frac{\gamma_at}{\lambda}\right)^2\right]+O\left(\frac{1}{\lambda}\right)^2,
		\nonumber \\
		\rho_{13}\approx&\frac{4\sqrt{2}}{\lambda}\rho_{24}(0)\exp\left[\frac{4 (\sqrt{2}-4)\gamma_at}{\lambda}+(2\sqrt{2}-3)\gamma_at+O\left(\frac{\gamma_at}{\lambda}\right)^2\right] +O\left(\frac{1}{\lambda}\right)^2,
	\label{seq10}
\end{align}
\endgroup
where we have expanded the elements of the density matrix to the first order of $\frac{1}{\lambda}$ and $\frac{\gamma_a t}{\lambda}$. Similar to other autonomous error correction schemes~\cite{Wang2022Apr, Lihm2018Jul}, the error correction efficiency based on the RL code also depends on the evaluation time $\gamma_{a}t$. If $\gamma_{a}t$ is too large, the system has decayed into the ground state ($\rho_{00}(t)\approx 1$), i.e., the capability for AQEC is lost. Therefore, we limit $\gamma_{a} t$ to an appropriate scale $\gamma_at\leq 0.6$. If $\lambda\gg 24$ (i.e., $\lambda\rightarrow\infty$), we can approximate the density matrix to order zero in the parameters $\frac{1}{\lambda}$ and $\frac{\gamma_a t}{\lambda}$
\begin{equation}
\rho_a(t)\approx\left(\begin{array}{ccccc}
	0 & 0 & 0 & 0 & 0\\
	0 & 0 & 0 & 0 & 0\\
	0 & 0 & \rho_{22}(0) & 0 & \rho_{24}(0)\exp(-u\gamma_{a}t)\\
	0 & 0 & 0 & 0 & 0\\
	0 & 0 & \rho_{42}(0)\exp(-u\gamma_{a}t) & 0 & \rho_{44}(0)
\end{array}\right), ~~u=3-2\sqrt{2}\approx 0.17. 
\label{seq11} 
\end{equation}
With this, we can derive the approximate mean fidelity of the RL code space  
\begin{equation}
	\begin{split}
	\bar{F}(t)&= \frac{1}{4\pi}\int^{2\pi}_{0}\int^{\pi}_{0}\left[\rho^2_{22}(0)+\rho^2_{44}(0)+2\vert\rho_{24}(0)\vert^2\exp({-u\gamma_{a}t})\right]\sin\theta~\mathrm{d}\theta\mathrm{d}\phi \\
	&=\frac{1}{2}\int^{\pi}_{0}\left\lbrace1+2\sin^2\left(\frac{\theta}{2}\right)\cos^2\left(\frac{\theta}{2}\right)\left[\exp({-u\gamma_{a}t})-1\right]\right\rbrace\sin\theta~\mathrm{d}\theta \\
	&=\frac{2}{3}+\frac{1}{3}\exp({-u\gamma_{a}t}),
	\end{split}
\label{seq12} 
\end{equation}
which can be further simplified as $\bar{F}(t)=1-\frac{1}{3}u\gamma_{a}t$ due to the condition $u\gamma_{a}t\ll 1$. 
\begin{figure}[bt]
	\includegraphics[width=10in]{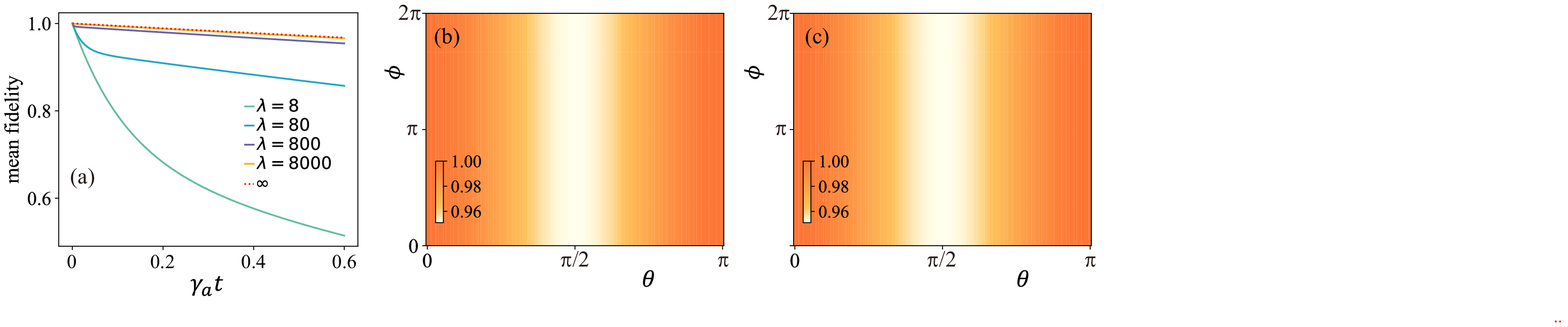}
	\caption{(a) Evolution of the mean fidelity for different choices of $\lambda$. (b) The analytic fidelity $F(\rho_a(\theta,\phi,0),\rho_a(\theta,\phi,t))$ (i.e., Eq.(\ref{seq13})) as function of the Bloch angles $\theta$ and $\phi$.  (c) The numerical fidelity $F(\rho_a(\theta,\phi,0),\rho_a(\theta,\phi,t))$ (i.e., numerically simulating the master equation Eq.(\ref{seq8}) with $\lambda=50, 000$) as function of the Bloch angles $\theta$ and $\phi$. Other parameters are $\gamma_at=0.6$.} 
	\label{sfig1}
\end{figure}
Moreover, the fidelity between an individual initial state $\rho_a(\theta,\phi,0)$ and the corresponding evolved state $\rho_a(\theta,\phi,t)$ is 
 \begin{equation}
F(\rho_a(\theta,\phi,0),\rho_a(\theta,\phi,t))=\operatorname{Tr}[\left|\psi_{\theta \phi}\rangle\langle \psi_{\theta \phi}\right\vert\rho(\theta,\phi,t)]=1+2\sin^2\left(\frac{\theta}{2}\right)\cos^2\left(\frac{\theta}{2}\right)\left[\exp({-u\gamma_{a}t})-1\right],
 \label{seq13} 
 \end{equation}
which depends on the Bloch angle $\theta$, but not on the Bloch angle $\phi$ for $\lambda \rightarrow \infty$. If the angle $\theta$ is equal to 0 and $\pi$, the fidelity is close to the maximum value of 1; if it is $\frac{\pi}{2}$, then the fidelity is the lowest (about 0.95), but still well above the break-even point (about 0.84).
For a comparison between the numerical and analytical results, we simulate the time-dependent mean fidelity with different $\lambda$ in Fig.~\ref{sfig1}(a). Our results demonstrate that the mean fidelity increases with increasing $\lambda$. Moreover, the numerical mean fidelity (i.e., simulating the master Eq.~(\ref{seq8}) with $\lambda=8000$) well agrees with the analytical solution Eq.~(\ref{seq12}) (i.e., $\lambda=\infty$). Similarly, we get very good agreement between analytical prediction and numerical evaluation for individual state fidelities, as shown in Figs.~\ref{sfig1}(b) and (c). Numerical fidelities are obtained by simulating the master Eq.~(\ref{seq8}) with $\lambda=50, 000$. This shows that our analytical results are reliable for $\lambda \gg 24$.

\section{Error-correctable fidelity}
\begin{figure}[bt]
	\includegraphics[width=5 in]{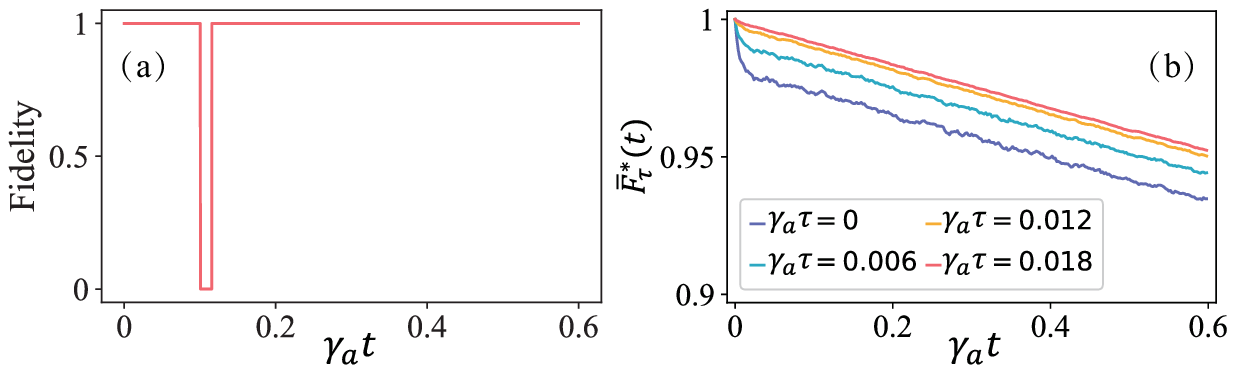}
	\caption{Performance of the RL code in terms of the error-correctable fidelity. (a) Fidelity  $\langle 2\vert \rho(t)\vert 2\rangle$ for a random quantum trajectory with the initial state $\vert 2\rangle$.	(b) Trajectory-wise, temporally coarse-grained fidelity $\bar{F}^*_{\tau}(t)$ , averaged over the code space and over $10^4$ quantum trajectories, with $ \gamma_a\tau=0,~0.006,~0.012,\text{and}~0.018$; other parameters are the same as in Fig.~\ref{sfig1q}. For sufficiently large $\tau$, the initial, delay-induced dip is removed. The resulting error-correctable fidelity reflects the irreversible fidelity loss.} 
	\label{SFigure_ad}
\end{figure}
As a consequence of the finite rate of the engineered jump operator in AQEC, there is a delay between the occurrence of an error and the onset of the recovery jump. This becomes transparent if one unravels the evolution under the master equation (\ref{seq7}) in terms of individual quantum trajectories. In~Fig.~\ref{SFigure_ad}(a) we demonstrate this with a random quantum trajectory for the initial state $\vert 2 \rangle$. Note that the fidelity drops to zero when an error occurs, and recovers to close to unity once the delayed recovery jump occurs (Let us clarify that other initial states, specifically, superpositions of the code words, would recover to fidelity values close to unity but lower, reflecting the approximate nature of the AQEC). Therefore, this random delay reduces the average fidelity when averaged over many trajectories, resulting in an initial dip as observed in Fig.~\ref{SFigure_ad}(b) for $\gamma_a \tau = 0$. However, the associated fidelity loss is, in principle, recoverable, as the information, while stored in the error space, is not lost. It is therefore instructive to distill the irreversible part of the fidelity loss, caused by the occurrence of multiple errors before correction and the imperfect state recovery under approximate AQEC. This can be achieved by using a temporally coarse-grained redefinition of the fidelity~\cite{Sarma2013Mar}:
\begin{equation}\label{horizon_fidelity}
F^*_{\tau}(t)=\max_{t^*\in[t,~t+\tau]}F(t^*) .
\end{equation}
If the coarse-graining parameter $\tau$ is chosen sufficiently large, i.e., on the order of the average time delay, then the fidelity (\ref{horizon_fidelity}) ignores the delay-induced fidelity loss, reflecting the error-correctable fidelity. In Fig.~\ref{SFigure_ad}(b) we demonstrate this with the RL code for different choices of $\tau$. As expected, we find that the initial dip of the fidelity is removed with increasing $\tau$.

While the error-correctable fidelity provides valuable insight about the in-principle achievable performance of the autonomously corrected RL code, the standard fidelity remains the operationally relevant one. This is because, in the absence of monitoring, that is, if no additional information is gained about whether the state resides in the code space or in the error space, gate operations must be conducted under the (due to the delayed-error-correction process not necessarily correct) assumption that, at the time of the gate operation, the state resides in the code space.


\section{Naive error correction operator}
\begin{figure}[hbt]
	\includegraphics[width=6.5in]{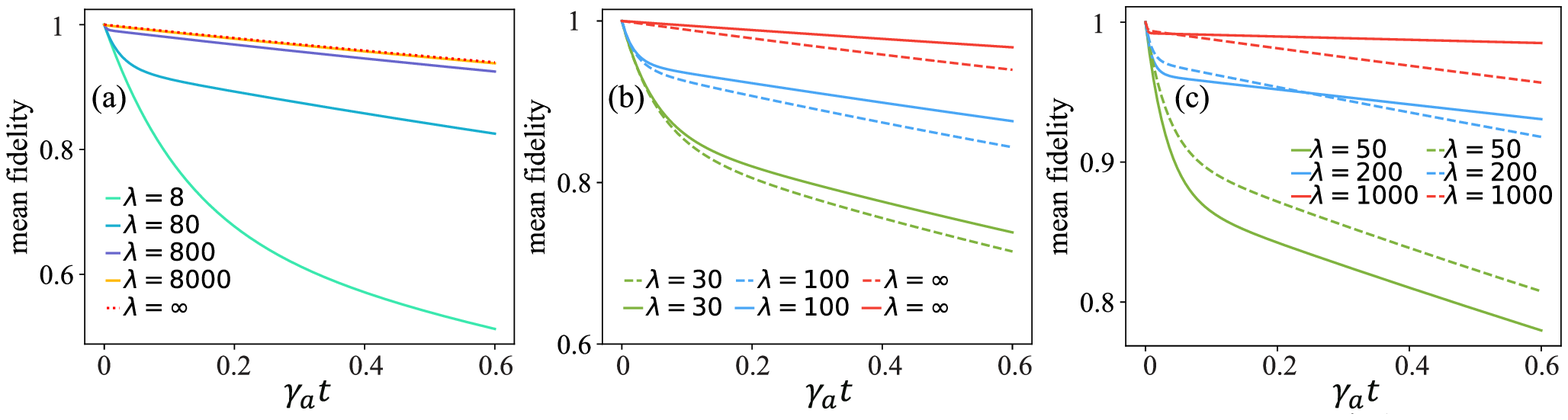}
	\caption{(a) Mean fidelity versus time for different $\lambda$ (the red dotted line describes the analytical result); (b) Comparison of the error correction performance of the error correction operator Eq.~(\ref{seq15}) (dashed line) and the RL error correction operator from the main text (solid line);  (c) Evolution of the mean fidelity resulting from the original master equation Eq. (\ref{seq8}) (dashed line) or from the modified master equation ($a \rightarrow a+a_1$) (solid line) for different $\lambda$.} 
	\label{sfig2}
\end{figure}
Let us consider another jump operator that may appear to counteract well the single-photon loss. When a single-photon loss occurs, the encoded state is changed to the error state
\begin{equation}
	a(\cos\frac{\theta}{2}\vert 2\rangle+e^{i \phi}\sin\frac{\theta}{2}\vert 4\rangle) ~~\longrightarrow~~ \cos\frac{\theta}{2}\sqrt{2}\vert 1\rangle+\sqrt{4}e^{i \phi}\sin\frac{\theta}{2}\vert 3\rangle.
	\label{seq14} 
\end{equation} 
To steer the error state back to the code space undisturbed, one may tentatively use the following error correction operator, which is different from the one in the main text:
\begin{equation}
	\begin{aligned}
		L_{\text{o}}=\sqrt{2}\vert 2\rangle\langle 1\vert+\vert 4\rangle\langle 3\vert, ~~~
		L_{\text{eng}}=\frac{L_{\text{o}}}{\sqrt{\operatorname{Tr}[L^{\dagger}_{\text{o}}L_{\text{o}}]}}.
	\end{aligned}
	\label{seq15}
\end{equation}
This jump operator revises the error state as
\begin{equation}
	L_{\text{eng}}\left(\cos\frac{\theta}{2}\sqrt{2}\vert 1\rangle+\sqrt{4}e^{i \phi}\sin\frac{\theta}{2}\vert 3\rangle\right)~~\longrightarrow~~ \frac{2}{\sqrt{\text{Tr}[L^{\dagger}_{\text{o}}L_{\text{o}}]}}\left(\cos\frac{\theta}{2}\vert 2\rangle+e^{i \phi}\sin\frac{\theta}{2}\vert 4\rangle\right),
	\label{seq16} 
\end{equation}  
that is, the encoded information remains undisturbed. To further investigate the efficiency of this error correction operator, we substitute it into the master Eq.~(\ref{seq8}) and restrict the system again to the first four photons, which yields the equations
\begin{equation}
\begin{aligned}
	\frac{1}{\gamma_a}\frac{d\rho_{22}}{dt} & =\frac{1}{2}(6\rho_{33}-4\rho_{22}+\frac{4}{3}\lambda\rho_{11}), ~~~~~~~~~	\frac{1}{\gamma_a}\frac{d\rho_{24}}{dt}  =\frac{1}{2}(-6\rho_{24}+\frac{2\sqrt{2}}{3}\lambda\rho_{13}),\\ 
	\frac{1}{\gamma_a}\frac{d\rho_{44}}{dt} & =\frac{1}{2}(-8\rho_{44}+\frac{2}{3}\lambda\rho_{33}), ~~~~~~~~~~~~~~~~ \frac{1}{\gamma_a}\frac{d\rho_{42}}{dt}  =\frac{1}{2}(-6\rho_{42}+\frac{2\sqrt{2}}{3}\lambda\rho_{31}),\\
	\frac{1}{\gamma_a}\frac{d\rho_{11}}{dt} & =\frac{1}{2}(4\rho_{22}-2\rho_{11}-\frac{4}{3}\lambda\text{\ensuremath{\rho_{11}}}),~~~~~~~~~
		\frac{1}{\gamma_a}\frac{d\rho_{13}}{dt}  =\frac{1}{2}(2\sqrt{8}\rho_{24}-4\rho_{13}-\lambda\text{\ensuremath{\rho_{13}}}),\\
	\frac{1}{\gamma_a}\frac{d\rho_{33}}{dt} & =\frac{1}{2}(8\rho_{44}-6\rho_{33}-\frac{2}{3}\lambda\text{\ensuremath{\rho_{33}}}),~~~~~~~~~  
	\frac{1}{\gamma_a}\frac{d\rho_{31}}{dt}  =\frac{1}{2}(2\sqrt{8}\rho_{42}-4\rho_{31}-\lambda\text{\ensuremath{\rho_{31}}}).
\end{aligned}
\label{seq17} 
\end{equation}
In the limit $\lambda\rightarrow\infty$, we obtain an approximate density matrix, which has the same form as Eq.(\ref{seq11}), except for that $u=\frac{1}{3}$. Correspondingly, the expressions for the fidelities $F(\vert\psi_{\theta \phi}\rangle\langle\psi_{\theta \phi}\vert,\rho_a(\theta,\phi,t))$ and $\bar{F}(t)$ remain the same, cf.~Eqs.~(\ref{seq12}) and (\ref{seq13}): the only difference is $u=\frac{1}{3}>3-2\sqrt{2}$. In Fig.~\ref{sfig2}(a), we show the fidelity over time for different $\lambda$. We find that the analytical results coincide well with the numerical results for large enough $\lambda$. In Fig.~\ref{sfig2}(b), we see that the efficiency of the RL error correction operator from the main text is higher than the error correction operator Eq.~(\ref{seq15}). Why is this? We can provide a physical reason: although the error correction operator Eq.~(\ref{seq15}) ensures that the information is not disturbed when the jumps occur, the time evolution in between the jumps, captured by the non-Hermitian terms in the master equation, acts detrimentally on the encoded information (i.e., $\lambda\langle 1\vert F^{\dagger}_{\text{eng}}F_{\text{eng}}\vert 1\rangle\gg \lambda\langle 3\vert F^{\dagger}_{\text{eng}}F_{\text{eng}}\vert 3\rangle$), resulting in an overall reduced performance compared to the RL error correction operator of the main text.

\section{Analysis of the Knill-Laflamme condition}

The mean fidelity of the RL code, Eq.~(\ref{seq12}) remains slightly below one because the code words have different mean photon numbers: the Knill-Laflamme (KL) condition is only partially satisfied \cite{PhysRevA.55.900}. To understand the consequences of this better, let us modify the master equation by replacing the single-photon loss jump operator $a$ by $a+a_1$, where $a_1=(2-\sqrt{2})\vert 1\rangle\langle 2\vert$ (i.e., $\mathcal{D}[a]\rightarrow\mathcal{D}[a+a_1]$). The modified jump operator now satisfies the condition $\langle 0_L\vert(a+a_1)^{\dagger}(a+a_1)\vert0_L\rangle=\langle 1_L\vert(a+a_1)^{\dagger}(a+a_1)\vert1_L\rangle$ for the RL code, that is, the KL condition is satisfied. Therefore, we can completely correct the error $a+a_1$ under the conditions $\frac{8g^2}{\gamma_a\gamma_b}\rightarrow\infty$, $\frac{1}{\lambda}\ll1$, and $\frac{\gamma_at}{\lambda}\ll1$. Next, we show that the mean fidelity can indeed reach unity under the 0-order approximation of the parameters $\frac{1}{\lambda}$ and $\frac{\gamma_{a}t}{\lambda}$.
To this end, let us again evaluate the corresponding master equation for the system  up to four photons,
\begin{equation}
\begin{split}
		\frac{1}{\gamma_a}\frac{d\rho_{22}}{dt}  &=\frac{1}{2}(6\rho_{33}-8\rho_{22}+\lambda\rho_{11}), ~~~~~~ ~~~~  \frac{1}{\gamma_a}\frac{d\rho_{24}}{dt}  =\frac{1}{2}(-8\rho_{24}+\lambda\rho_{13}), \\
	\frac{1}{\gamma_a}\frac{d\rho_{44}}{dt} &=	\frac{1}{2}(-8\rho_{44}+\lambda\rho_{33}), ~~~ ~~~ ~~~~~~~~~ ~~ \frac{1}{\gamma_a}\frac{d\rho_{42}}{dt}  	=\frac{1}{2}(-8\rho_{42}+\lambda\rho_{31}),
	\\
		\frac{1}{\gamma_a}\frac{d\rho_{11}}{dt}  	&=	\frac{1}{2}(8\rho_{22}-2\rho_{11}-\lambda\text{\ensuremath{\rho_{11}}}),~~~~~~~~~~ 	\frac{1}{\gamma_a}\frac{d\rho_{13}}{dt}  =\frac{1}{2}(8\rho_{24}-4\rho_{13}-\lambda\text{\ensuremath{\rho_{13}}}),
		\\
		\frac{1}{\gamma_a}\frac{d\rho_{33}}{dt} &=	\frac{1}{2}(8\rho_{44}-6\rho_{33}-\lambda\text{\ensuremath{\rho_{33}}}), ~~~~~~~~~~  \frac{1}{\gamma_a}\frac{d\rho_{31}}{dt}  =\frac{1}{2}(8\rho_{42}-4\rho_{31}-\lambda\text{\ensuremath{\rho_{31}}}) .
\end{split}
\label{seq19}
\end{equation}
We approximately solve this equation by expanding to first order in the parameters $\lambda\gg 1$ and $\lambda\gg \gamma_a t$,
\begingroup
\allowdisplaybreaks
\begin{align}
		\rho_{00}(t)\approx& 1
		-\left(\frac{3}{\lambda}+\frac{3}{2}\right)\rho_{44}(0)\left\lbrace\exp\left[-\frac{8\gamma_{a}t}{\lambda}+O\left(\frac{\gamma_{a}t}{\lambda}\right)^2 \right]-\exp\left[-\frac{24\gamma_{a}t}{\lambda}+O\left(\frac{\gamma_{a}t}{\lambda}\right)^2\right]\right\rbrace \nonumber
		\\
		 &-\rho_{44}(0)\exp\left[-\frac{24\gamma_{a}t}{\lambda}+O\left(\frac{\gamma_{a}t}{\lambda}\right)^2 \right]-\rho_{22}(0)\exp\left[-\frac{8\gamma_{a}t}{\lambda}+O\left(\frac{\gamma_{a}t}{\lambda}\right)^2 \right]+O\left(\frac{1}{\lambda}\right)^2,\nonumber
		\\
		 \rho_{11}(t)\approx&\frac{12}{\lambda}\rho_{44}(0)\left\lbrace\exp\left[-\frac{8\gamma_{a}t}{\lambda}+O\left(\frac{\gamma_{a}t}{\lambda}\right)^2 \right]-\exp\left[-\frac{24\gamma_{a}t}{\lambda}+O\left(\frac{\gamma_{a}t}{\lambda}\right)^2\right]\right\rbrace\nonumber
		 \\
		  &+\frac{8}{\lambda}\rho_{22}(0)\exp\left[-\frac{8\gamma_{a}t}{\lambda}+O\left(\frac{\gamma_{a}t}{\lambda}\right)^2 \right]+O\left(\frac{1}{\lambda}\right)^2,\nonumber
		 \\
		\rho_{22}(t)\approx&\left(\frac{3}{2}-\frac{9}{\lambda}\right)\rho_{44}(0)\left\lbrace\exp\left[-\frac{8\gamma_{a}t}{\lambda}+O\left(\frac{\gamma_{a}t}{\lambda}\right)^2 \right]-\exp\left[-\frac{24\gamma_{a}t}{\lambda}+O\left(\frac{\gamma_{a}t}{\lambda}\right)^2\right]\right\rbrace\nonumber
		 \\  &+\left(1-\frac{8\gamma_{a}t}{\lambda}\right)\rho_{22}(0)\exp\left[-\frac{8\gamma_{a}t}{\lambda}+O\left(\frac{\gamma_{a}t}{\lambda}\right)^2 \right]+O\left(\frac{1}{\lambda}\right)^2, \nonumber
		\\	
		\rho_{33}(t)\approx&\frac{8}{\lambda}\rho_{44}(0)\exp\left[-\frac{24\gamma_{a}t}{\lambda}+O\left(\frac{\gamma_at}{\lambda}\right)^2 \right]+O\left(\frac{1}{\lambda}\right)^2,\nonumber 
			\\
			\rho_{44}(t)\approx&\left(1-\frac{8}{\lambda}\right)\rho_{44}(0)\exp\left[-\frac{24\gamma_{a}t}{\lambda}+O\left(\frac{\gamma_{a}t}{\lambda}\right)^2 \right]+O\left(\frac{1}{\lambda}\right)^2, \nonumber
			\\
			\rho_{13}(t)\approx& \frac{8}{\lambda}\rho_{24}(0)\exp\left[-\frac{16}{\lambda}\gamma_{a}t+O\left(\frac{\gamma_{a}t}{\lambda}\right)^2 \right]+O\left(\frac{1}{\lambda}\right)^2,\nonumber
			 \\
		\rho_{24}(t)\approx&\left(1-\frac{8}{\lambda}\right)\rho_{24}(0)\exp\left[-\frac{16}{\lambda}\gamma_{a}t+O\left(\frac{\gamma_{a}t}{\lambda}\right)^2 \right]+O\left(\frac{1}{\lambda}\right)^2.
		\label{seq20}
\end{align}
\endgroup
The density matrix is approximately equal to the initial state $\rho_a(0)$ for $\lambda\rightarrow\infty$, so the mean fidelity is approximately equal to unity.
This implies that we can fully correct the single-photon loss by adding the small correction term $a_1$. In Fig.~\ref{sfig2}(c), we compare the error correction performance of the modified master equation ($a \rightarrow a+a_1$) and the original master equation (\ref{seq8}) for different values of $\lambda$. With increasing $\lambda$, the modified master equation displays a better error correction ability. When $\lambda$ is small, the mean fidelity is lower due to $\langle 2\vert (a+a_1)^{\dagger}(a+a_1)\vert2\rangle>\langle 2\vert a^{\dagger}a\vert2\rangle$.

\begin{figure}[bt]
	\includegraphics[width=3.6in]{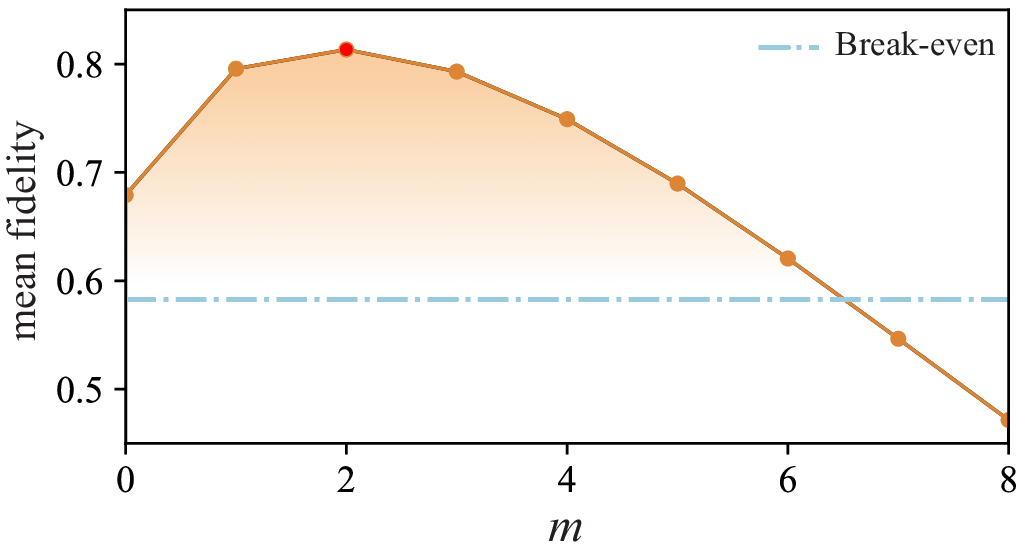} \caption{ Simulating the mean fidelity at $t=150~\mu$s for different $m$ with the master equation (5) of the main text. Other parameters are $\gamma_a=0.02$~MHz, $g=8$~MHz, and $\gamma_b/g=2.5$.} 
	\label{fig3}
\end{figure}

Finally, let us discuss the optimality of the RL code compared to codewords that are shifted in Fock space, i.e., have the general form $\vert m\rangle$, $\vert m+2\rangle$ (for example, $\vert 1\rangle,~\vert 3\rangle$ or $\vert8\rangle, ~\vert 10\rangle$). While the code space better satisfies the KL condition with increasing $m$, the mean probability of single-photon jumps also increases,
\begin{equation}
	\begin{split}
		\bar{P}_{\text{er}}\propto \frac{1}{4\pi}\int_{\Omega}\langle \psi_{\theta\phi}\vert \gamma_a a^{\dagger}a\vert \psi_{\theta\phi}\rangle~\mathrm{d}\Omega
		=(m+1)\gamma_a.
	\end{split}
	\label{eq8}
\end{equation}
Therefore, high excitation numbers require large $g/\gamma_{a}$ for the jump probability of $L_{\text{eng}}$ to be sufficiently large, which increases the difficulty of the experiment. Here we limit the range of the cooperativity $C\leq 160$ to ensure experiment-friendly AQEC. Under this parameter condition, we simulate the mean fidelity for the code word $\vert m\rangle$, $\vert m+2\rangle$ in Fig.~\ref{fig3}. We find that the mean fidelity of the RL code is optimal compared to its shifted versions. It is worth noting that $m=0$ is an invalid code space due to $F(\frac{\pi}{2},\phi,t)<\bar{F}_{\text{be}}$. 

\section{Setting the coupling between the qubit \& the encoding mode plus the auxiliary mode}
Here we discuss in more detail how to realize our approximate AQEC scheme in an actual physical system. We consider a quantum system consisting of an encoding mode, an auxiliary qubit, and an auxiliary mode. The system Hamiltonian is given by
\begin{equation}
	H=\omega_aa^{\dagger}a+\frac{\omega_b}{2}\sigma_{z}+\omega_cc^{\dagger}c+\frac{\chi}{2}a^{\dagger}a\sigma_{z}+f(t)(a+a^{\dagger})\sigma_{x}+g_c(t)(c^{\dagger}+c)\sigma_{x},
	\label{seq22}
\end{equation}
where $\chi$ is the dissipative coupling coefficient, $g(t)$ and $f(t)$ are two time-dependent control fields, and $\omega_a$, $\omega_b$, $\omega_c$ are the resonant frequencies of the encoding mode, the qubit, and the auxiliary mode, respectively. 
The system dynamics are then described by the master equation
\begin{equation}
	\frac{d\rho}{dt}=-i[H, \rho]+\frac{\gamma_{a1}}{2}\mathcal{D}[a]+\frac{\gamma_{b1}}{2}\mathcal{D}[\sigma_-]+\frac{\gamma_{c1}}{2}\mathcal{D}[c],
	\label{seq23}
\end{equation}
where $\gamma_{i1}$ ($i=a,b,c$) are the corresponding decay rates. The Hamiltonian~(\ref{seq22}) can be expanded with the eigenstates $\vert E_{N,M,\pm}\rangle$ of the Hamiltonian $H_0=\omega_aa^{\dagger}a+\frac{\omega_b}{2}\sigma_{z}+\omega_cc^{\dagger}c+\frac{\chi}{2}a^{\dagger}a\sigma_{z}$. Therefore, the Hamiltonian $H$ can be rewritten as
\begin{equation}
	\begin{split}
		H\!=\!&\sum_{N,M}E_{N,M}\vert E_{N,M,\pm}\rangle\langle E_{N,M,\pm}\vert+f(t)\sqrt{N+1}(\vert E_{N+1,M,+}\rangle\langle E_{N,M,-}\vert+\vert E_{N+1,M,-}\rangle\langle E_{N,M,+}\vert+h.c.) \\
		&+g_c(t)\sqrt{M+1}(\vert E_{N,M+1,+}\rangle\langle E_{N,M,-}\vert+\vert E_{N,M+1,-}\rangle\langle E_{N,M,+}\vert+h.c.),
	\end{split}	
\end{equation}
where $E_{N,M,\pm}=N\omega_a+M\omega_c\pm(\frac{\omega_b}{2}+\frac{N\chi}{2})$ is the eigenvalue of the Hamiltonian $H_0$. After switching to a rotating frame by applying the unitary operator $U=\exp(-iH_0t)$, the Hamiltonian can be written as
\begin{equation}
\begin{aligned}
H_I\!=\!\sum_{N,M}f(t)\sqrt{N+1}(\vert E_{N+1,M,+}\rangle\langle E_{N,M,-}\vert e^{i(E_{N+1,M,+}-E_{N,M,-})t}+\vert E_{N+1,M,-}\rangle\langle  E_{N,M,+}\vert e^{i(E_{N+1,M,-}-E_{N,M,+})t})\\
\!+\!g_c(t)\sqrt{M+1}(\vert E_{N,M+1,+}\rangle\langle E_{N,M,-}\vert e^{i(E_{N,M+1,+}-E_{N,M,-})t}\!+\!\vert E_{N,M+1,-}\rangle\langle E_{N,M,+}\vert e^{i(E_{N,M+1,-}-E_{N,M,+})t})+h.c. ~.
\end{aligned}
\label{seq24}
\end{equation}
\begin{figure}[bt]
	\includegraphics[width=3.5in]{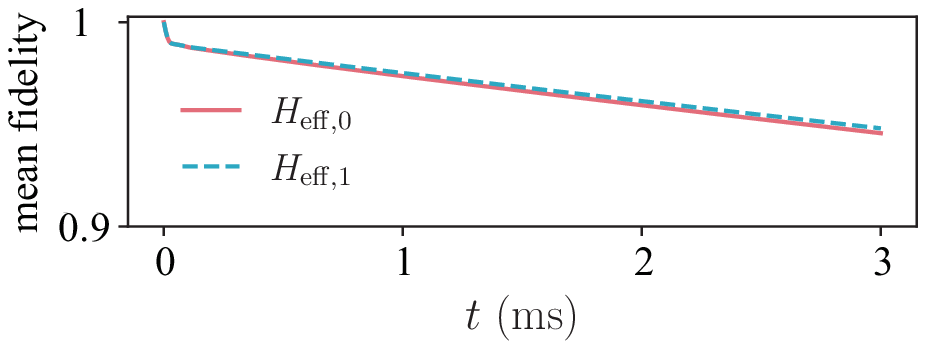} \caption{Evolution of the mean fidelity up to $t=3$ ms for the effective Hamiltonians $H_{\text{eff},0}$ and $H_{\text{eff},1}$, with the parameters $\alpha_0/2\pi=0.05$ MHz, $\alpha_1/2\pi=0.07$ MHz, $\gamma_{a1}/2\pi=0.2$ kHz, $\gamma_{b1}/2\pi=2$ kHz, and $\gamma_{c1}/2\pi=0.12$ MHz.    
	} 
	\label{sfig5}
\end{figure}
Here we assume for simplicity that the mode $c$ and the qubit are resonant, $\omega_b=\omega_c=\omega$, and that the frequencies satisfy $\vert\omega_a+\omega\vert,\vert\omega_a-\omega\vert\gg\chi\gg g_c(t), \vert f(t)\vert$. We can then use the control fields  
\begin{equation}
	\begin{split}
	&f(t)=\frac{2\alpha_0}{\sqrt{2}}\cos\left[\left(E_{2,M,+}-E_{1,M,-}\right)t\right]+\frac{2\alpha_0}{\sqrt{4}}\cos\left[\left(E_{4,M,+}-E_{3,M,-}\right)t\right], \\
	&g(t)=2\alpha_1\cos\left[(E_{2,M,+}-E_{2,M+1,-})t\right]+2\alpha_1\cos\left[(E_{4,M,+}-E_{4,M+1,-})t\right],
	\end{split}
	\label{seq25}
\end{equation}
to select effective transitions.
The control fields must satisfy the condition $\vert f(t)\vert\leq\vert g(t)\vert$ to  rapidly transfer the energy from the auxiliary qubit to the auxiliary mode. After performing a rotating-wave approximation, we obtain the following effective Hamiltonian
\begin{equation}
	\begin{split}
		H_{\text{eff,0}}\approx\alpha_0 (L_o\sigma_+ + L_o^{\dagger}\sigma_-)+\sum_{N=2,4}\alpha_1\vert N\rangle\langle N\vert(c^{\dagger}\sigma_-+c\sigma_+), 
	\end{split}
	\label{seq26}
\end{equation}
which can be further approximated to obtain the simpler expression
\begin{equation}
	\begin{split}
		H_{\text{eff,1}}\approx\alpha_0 (L_o\sigma_+ + L_o^{\dagger}\sigma_-)+\alpha_1(c^{\dagger}\sigma_-+c\sigma_+), 
	\end{split}
	\label{seq27}
\end{equation}
due to the fast energy exchange mainly appearing in the code space rather than the error space for the approximate AQEC.

 We numerically evaluate the mean fidelity under the two Hamiltonians $H_{\text{eff},0}$ and $H_{\text{eff},1}$ in Fig.~(\ref{sfig5}), which demonstrates that the above approximation is appropriate. The frame rotation does not affect the decay $D[\sigma_-]$ of the auxiliary mode. Although $D[a]$ and $D[\sigma_-]$ are modified in the rotating frame, the effects of $D[a]$ and $D[\sigma_-]$ remain approximately the same as in the original frame. Finally, we obtain the effective Hamiltonian
 \begin{equation}
 H_{\text{eff}}\propto L_{\text{eng}}\sigma_+ + L^{\dagger}_{\text{eng}}\sigma_-,
 \end{equation}
  (i.e., the QEC Hamiltonian of the main text) by adiabatically eliminating the high-decay mode $c$ \cite{Liu2014May}. Therefore, the dynamics described by Eq.(\ref{seq23}) are effectively equivalent to the master Eq.~(5) of the main text.

%
